\documentclass[11pt]{article}
\usepackage{amsmath,amsxtra}
\usepackage{amscd}
\usepackage{graphicx}
\usepackage{latexsym}
\usepackage{amsmath,amsthm,amsfonts,amssymb,cite}%,showlabels}
\usepackage{latexsym}
\usepackage{amsmath}

\newtheorem{remark}{Remark}
\numberwithin{equation}{section}
\begin{document}
\hoffset = -2.4truecm \voffset = -2truecm
\renewcommand{\baselinestretch}{1.2}
\newcommand{\mb}{\makebox[10cm]{}\\ }
\date{}
%%%%% DOCUMENT SPECIFIC DEFINITIONS

%  Theorems, Lemmas and the like, should be typeset in italic
\newtheorem{theorem}{Theorem}
\newtheorem{proposition}{Proposition}
\newtheorem{lemma}{Lemma}
\newtheorem{definition}{Definition}

%%%%% END DOCUMENT SPECIFIC DEFINITIONS
%\renewcommand{\square}{\hfill$\Box$\vspace{2ex}}
%\renewcommand{\Theta}{\Ta}

\title{Two hierarchies of new generalized multicomponent AKNS-type soliton equations }
\author{Chun-Xia Li$^{1}$\footnote{trisha\_li2001@163.com}, Shou-Feng Shen$^{2}$, Wen-Xiu Ma$^{3}$ and Shui-Meng Yu$^{4}$\\
$^{1}$School of Mathematical Sciences,
Capital Normal University, \\
Beijing 100048, PR China\\
$^{2}$Department of Applied Mathematics, Zhejiang University of Technology,\\
Hangzhou 310023, PR China\\
$^{3}$Department of Mathematics and Statistics, University of South Florida,\\
Tampa, FL 33620-5700, USA\\
$^{4}$ School of Sciences, Jiangnan University, Wuxi 214122, PR China
}
\date{}
\maketitle
\begin{abstract}
Two multicomponent generalizations of the AKNS-type spectral problems associated with $sl(2,\mathbb{R})$ and $so(3,\mathbb{R})$ are introduced and the corresponding two hierarchies of generalized multicomponent AKNS-type soliton equations are presented by the standard procedure, respectively. By virtue of the trace identity, bi-Hamiltonian structures which lead to a common recursion operator are established for each of the two resulting soliton hierarchies. And thus the Liouville integrability is shown for all systems in each of the two new generalized soliton hierarchies, seperately.
\end{abstract}

\indent\indent\indent\indent\indent  Recursion operators, Bi-Hamiltonian structures, Liouville integrability

\section{Introduction}
In literature, there has been a lot of work on how to generate soliton hierarchies from matrix spectral problems or Lax pairs by standard procedure\cite{AC,NM,CH,DS,ASF}. Among these examples, there are the celebrated Korteweg-de Vries(KdV) hierarchy, Ablowitz-Kaup-Newell-Segur(AKNS) hierarchy, Kaup-Newell(KN) hierarchy, Wadati-Konno-Ichikawa(WKI) hierarchy, the Dirac hierarchy and the Boiti-Pempinelli-Tu(BPT) hierarchy\cite{PDL,AKNS,KN,WKI,DR,BPT}. Soliton hierarchies generated from spectral problems often possess bi-Hamiltonian structures and are integrable in the sense of Liouville. The corresponding bi-Hamiltonian structures can be established by the variational identity \cite{MA4,MA5}, particularly by the trace identity when the underlying matrix loop algebra is semisimple \cite{TGZ}.

Recently, Ma etc. successfully proposed some new matrix spectral problems by generalizing the classical AKNS, KN and WKI spectral problems associated with the Lie algebra $sl(2,\mathbb{R})$ to the ones associated with the Lie algebra $so(3,\mathbb{R})$, derived their corresponding soliton hierarchies and established bi-Hamiltonian structures by standard procedure described in \cite{TGZ,MA1,MA2,MA3}.

The three-dimensional real special orthogonal Lie algebra $so(3,\mathbb{R})$ consists of $3\times 3$ skew-symmetric matrices. This Lie algebra is simple and has the basis
\begin{equation}
e_1=\begin{pmatrix}0&0&-1\\ 0&0&0\\ 1&0&0\end{pmatrix},\,e_2=\begin{pmatrix}0&0&0\\ 0&0&-1\\ 0&1&0\end{pmatrix},\,e_3=\begin{pmatrix}0&-1&0\\ 1&0&0\\ 0&0&0\end{pmatrix},
\end{equation}
whose commutator relations are
\begin{align*}
[e_1,e_2]=e_3,\, [e_2,e_3]=e_1,\, [e_3,e_1]=e_2.
\end{align*}
The derived algebra of $so(3,\mathbb{R})$ is $so(3,\mathbb{R})$ itself. The other real three-dimensional Lie algebra with a three-dimensional derived algebra is $sl(2,\mathbb{R})$ which has been widely used in soliton theory. In fact, assume that the Lie algebra $sl(2,\mathbb{R})$ has the basis
\begin{equation}
e_1'=\begin{pmatrix}1&0\\ 0&-1\end{pmatrix},\,e_2'=\begin{pmatrix}0&1\\ 0&0\end{pmatrix},\,e_3'=\begin{pmatrix}0&0\\ 1&0\end{pmatrix},
\end{equation}
it is not difficult to prove that the basis $e_1'$, $e_2'$ and $e_3'$ satisfy the following relations
\begin{align*}
[e_1',e_2']=2e_2',\,[e_3',e_1']=2e_3',\,[e_2',e_3']=e_1'.
\end{align*}

The classical AKNS spectral problem reads as
\begin{align}
\phi_x=(-\lambda e_1'+qe_2'+re_3')\phi.\label{AKNS}
\end{align}
In \cite{MA1}, Ma generalized the spectral problem \eqref{AKNS} to the one associated with $so(3,\mathbb{R})$:
\begin{align}\label{AKN1}
\phi_x=(-\lambda e_1+pe_2-qe_3)\phi.
\end{align}
In \cite{YZY}, Yan etc. proposed another generalization of \eqref{AKNS}:
\begin{align}
\phi_x=(-(\lambda+\mu qr) e_1'+qe_2'+re_3')\phi.\label{GAKNS}
\end{align}
Stimulated by Ma's and Yan's work, in \cite{LSMY}, we proposed the following spectral problem:
\begin{align}\label{AKNSS}
\phi_x=[-(\lambda+\beta (p^2+q^2) )e_1+pe_2-qe_3]\phi.
\end{align}

In this paper, we shall generalize the above two spectral problems \eqref{GAKNS} and \eqref{AKNSS} to their multicomponent counterparts, study their corresponding integrable soliton hierarchies and bi-Hamiltonian structures, and finally prove their Liouville integrability, respectively.

%In Yan's paper, the following generalized AKNS spectral problem was studied.
%\begin{align}
%\phi_x=\begin{pmatrix}
%-\lambda-\mu qr&q\\
%r&\lambda+\mu qr
%\end{pmatrix}\phi=[-(\lambda+\mu qr)e_1'+qe_2'+re_3']\phi
%\end{align}
%where $\mu$ is an arbitrary constant, $e_1'$, $e_2'$ and $e_3'$ are the basis of $sl(2,R)$ given by
%\begin{equation}
%e_1'=\begin{pmatrix}1&0\\ 0&-1\end{pmatrix},e_2'=\begin{pmatrix}0&1\\ 0&0\end{pmatrix},e_3'=\begin{pmatrix}0&0\\ 1&0\end{pmatrix}.
%\end{equation}
%It is easy to calculate that the basis $e_1'$, $e_2'$ and $e_3'$ satisfy the following relations
%\begin{align}
%[e_1',e_2']=2e_2',[e_3',e_1']=2e_3',[e_2',e_3']=e_1'.
%\end{align}

\section{A hierarchy of generalized multicomponent AKNS equations associated with $sl(m+1,\mathbb{R})$}
\subsection{A hierarchy of generalized multicomponent AKNS equations}
Let $m$ be an arbitrary natural number. We consider the following $(m+1)\times (m+1)$ matrix spectral problem which we call the generalized multicomponent AKNS spectral problem
\begin{align}\label{GAKNS1}
\phi_x=U(u,\lambda)\phi=
\begin{pmatrix}
-m(\lambda+\beta r)&q\\
p&(\lambda+\beta r)I_m
\end{pmatrix}\phi,
\end{align}
where $\lambda$ is a spectral parameter, $I_m$ is the $m\times m$ identity matrix and
\begin{align}
q&=(q_1,q_2,\cdots,q_m), \quad p=(p_1,p_2,\cdots,p_m)^T,\\
\phi&=(\phi_1,\phi_2,\cdots,\phi_{m+1}),\quad u=(q,p^T)^T,\quad  r=qp.
\end{align}

To derive the corresponding soliton hierarchy, let us first solve the stationary zero-curvature equation
\begin{align}
W_x=[U,W].\label{SZC1}
\end{align}
We assume the solution $W$ is given by
\begin{align}
 W=\begin{pmatrix}a&b\\c&d\end{pmatrix},
\end{align}
where $a$ is a scalar, $b^T$ and $c$ are $m$-dimensional column vectors, and $d$ is an $m\times m$ matrix. Therefore the stationary zero-curvature equation \eqref{SZC1} is equivalent to
\begin{equation}\left\{\begin{array}{l}
a_x=qc-bp,\\
d_x=pb-cq,\\
b_x=qd-aq-(m+1)(\lambda+\beta r)b,\\
c_x=pa-dp+(m+1)(\lambda+\beta r)c.
\end{array}\right.
\end{equation}
As a consequence, we can also derive that
\begin{align}
\mbox{tr}(d)_x=\mbox{tr}(d_x)=\mbox{tr}(pb-cq)=bp-qc=-a_x.
\end{align}

Let us seek a formal solution of the type
\begin{align}
W=\begin{pmatrix}a&b\\c&d\end{pmatrix}=\sum\limits_{k=0}^{\infty}W_k\lambda^{-k}=\sum\limits_{k=0}^{\infty}
\begin{pmatrix}a^{(k)}&b^{(k)}\\c^{(k)}&d^{(k)}\end{pmatrix}\lambda^{-k}
\end{align}
with $b^{(k)},c^{(k)}$ and $d^{(k)}$ being assumed to be
\begin{align}
b^{(k)}=(b_1^{(k)},b_2^{(k)},\cdots,b_m^{(k)}),\quad c^{(k)}=(c_1^{(k)},c_2^{(k)},\cdots,c_m^{(k)})^T,\quad d^{(k)}=(d_{ij}^{(k)})_{m\times m}.
\end{align}
%\begin{align}
%\sum\limits_{k\ge 0}a^{(k)}_{x}\lambda^{-k}&=\sum\limits_{k\ge 0}q c^{(k)}\lambda^{-k}-\sum\limits_{k\ge 0}b^{(k)} p\lambda^{-k},\\
%\sum\limits_{k\ge 0}b^{(k)}_{x}\lambda^{-k}&=\sum\limits_{k\ge 0}q d^{(k)}\lambda^{-k}-\sum\limits_{k\ge 0} a^{(k)}q\lambda^{-k}-(m+1)\beta r\sum\limits_{k\ge 0}b^{(k)}\lambda^{-k}-(m+1)\sum\limits_{k\ge 0}b^{(k)}\lambda^{-k+1},\\
%\sum\limits_{k\ge 0}c^{(k)}_{x}\lambda^{-k}&=\sum\limits_{k\ge 0}p a^{(k)}\lambda^{-k}-\sum\limits_{k\ge 0} d^{(k)}p\lambda^{-k}+(m+1)\beta r\sum\limits_{k\ge 0}c^{(k)}\lambda^{-k}+(m+1)\sum\limits_{k\ge 0}c^{(k)}\lambda^{-k+1},\\
%\sum\limits_{k\ge 0}d^{(k)}_{x}\lambda^{-k}&=\sum\limits_{k\ge 0}p b^{(k)}\lambda^{-k}-\sum\limits_{k\ge 0}c^{(k)} q\lambda^{-k}.
%\end{align}
%Taking the initial values
%\begin{align}
%a_0=1,b_0=c_0=0,
%\end{align}
Under these assumptions, we have the following recursion relation
\begin{equation}\left\{\begin{array}{l}
b^{(0)}=0,\quad c^{(0)}=0,\quad a^{(0)}_x=0,\quad d^{(0)}_x=0,\\
a^{(k)}_{x}=qc^{(k)}-b^{(k)}p,\quad d^{(k)}_x=pb^{(k)}-c^{(k)}q,\quad k\ge 0,\\
b^{(k+1)}=\frac{1}{m+1}[qd^{(k)}-a^{(k)}q-b^{(k)}_x-(m+1)\beta rb^{(k)}],\quad k\ge 0,\\
c^{(k+1)}=\frac{1}{m+1}[c^{(k)}_x-pa^{(k)}+d^{(k)}p-(m+1)\beta rc^{(k)}],\quad k\ge 0.\\
\end{array}\right.
\end{equation}
from which, we can obtain the recursion relation for $b^{(k)}$ and $c^{(k)}$:
\begin{align}
\begin{pmatrix}
c^{(k+1)}\\{b^{(k+1)}}^T
\end{pmatrix}=L\begin{pmatrix}c^{(k)}\\{b^{(k)}}^T\end{pmatrix},\quad k\ge 0
\end{align}
with $L$ given by
\begin{align}
L=\frac{1}{m+1}
\begin{pmatrix}
[\partial-\beta(m+1)r-\sum\limits_{i=1}^m p_i\partial^{-1}q_i]I_m-p\partial^{-1}q&p\partial^{-1}p^T+(p\partial^{-1}p^T)^T\\
-q^T\partial^{-1}q-(q^T\partial^{-1}q)^T&[-\partial-\beta(m+1)r+\sum\limits_{i=1}^mq_i\partial^{-1}p_i]I_m+q^T\partial^{-1}p^T
\end{pmatrix}.\end{align}
%\begin{align}
%L^\dagger=\frac{1}{m+1}\begin{pmatrix}
%[-\partial-\beta(m+1)r+\sum\limits_{i=1}^m q_i\partial^{-1}p_i]I_m+q^T\partial^{-1}p^T&q^T\partial^{-1}q+(q^T\partial^{-1}q)^T\\
%-p\partial^{-1}p^T-(p\partial^{-1}p^T)^T&[\partial-\beta(m+1)r-\sum\limits_{i=1}^mp_i\partial^{-1}q_i]I_m-p\partial^{-1}q
%\end{pmatrix}
%\end{align}

By choosing the initial values to be
\begin{align}
a^{(0)}=-m,\quad d^{(0)}=I_m,
\end{align}
and requiring that
\begin{align}
W_k|_{u=0}=0,\quad k\ge 1,
\end{align}
we can determine the sequence $\{a^{(k)},b^{(k)},c^{(k)},d^{(k)}|\,\,k\ge 1\}$ uniquely. The first few sets can be computed as
\begin{equation*}\left\{\begin{array}{l}
a^{(1)}=0,\quad b^{(1)}=q,\quad c^{(1)}=p,\quad d^{(1)}=(0)_{m\times m},\\
a^{(2)}=\frac{r}{m+1},\quad b^{(2)}=\frac{1}{m+1}[-q_x-(m+1)\beta rq],\\
c^{(2)}=\frac{1}{m+1}[p_x-(m+1)\beta rp],\quad d^{(2)}=\frac{1}{m+1}(-pq),\\
a^{(3)}=\frac{1}{(m+1)^2}[qp_x-q_xp-2(m+1)\beta r^2],\\
d^{(3)}=\frac{1}{(m+1)^2}[pq_x-p_xq+2(m+1)\beta rpq],\\
b^{(3)}=\frac{1}{(m+1)^2}\left\{q_{xx}-2rq+(m+1)\beta[r_xq+2rq_x+(m+1)\beta r^2q]\right\},\\
c^{(3)}=\frac{1}{(m+1)^2}\left\{p_{xx}-2pr-(m+1)\beta[r_xp+2rp_x-(m+1)\beta r^2p]\right\}.\end{array}\right.
\end{equation*}

Next, let us introduce the auxiliary problem
\begin{align}\label{AP1}
\phi_{t_n}=V^{(n)}\phi, \quad V^{(n)}&=(\lambda^nW)_++\Delta_n=\sum\limits_{j=0}^nW_j\lambda^{n-j}+\begin{pmatrix}e_n&0\\
0&h_n\end{pmatrix},\quad n\ge 0.
\end{align}
where $P_+$ denotes the polynomial part of $P$ in $\lambda$. The compatibility condition of \eqref{GAKNS1} and \eqref{AP1} generates the zero-curvature equation
\begin{align}
U_{t_n}-V^{(n)}_x+[U,V^{(n)}]=0,
\end{align}
which is equivalent to
\begin{equation}\left\{\begin{array}{l}
e_{n_x}=-m\beta r_{t_n},\quad h_{n_x}=\beta r_{t_n}I_m,\\
q_{t_n}=-(m+1)b^{(n+1)}-qh_n+e_nq,\\
p_{t_n}=(m+1)c^{(n+1)}-pe_n+h_np.\end{array}\right.
\end{equation}
By assuming $e_n=-mf_n$ and $h_n=f_nI_m$, we have
\begin{equation}\left\{\begin{array}{l}
{f_n}_x=\beta r_{t_n}=\beta(m+1)(qc^{(n+1)}-b^{(n+1)}p),\\
q_{t_n}=-(m+1)(b^{(n+1)}+f_nq),\\
p_{t_n}=(m+1)(c^{(n+1)}+f_np).\end{array}\right.
\end{equation}
Thus we can solve $f_n$ to get $f_n=\beta(m+1)a^{(n+1)}$. We finally obtain the generalized multicomponent AKNS hierarchy
\begin{equation}\left\{\begin{array}{l}
q_{t_n}=-(m+1)[b^{(n+1)}+\beta(m+1)\partial^{-1}(qc^{(n+1)}-b^{(n+1)}p)q],\\
p_{t_n}=(m+1)[c^{(n+1)}+\beta(m+1)\partial^{-1}(qc^{(n+1)}-b^{(n+1)}p)p],\end{array}\right.
\end{equation}
which can be rewritten as
\begin{align}\label{SH1}
u_{t_n}=\begin{pmatrix}q^T\\p
\end{pmatrix}_{t_n}=K_n=R\begin{pmatrix}c^{(n+1)}\\{b^{(n+1)}}^T\end{pmatrix},
\end{align}
with
\begin{align}
R=(m+1)
\begin{pmatrix}
-\beta(m+1)q^T\partial^{-1}q&-I_m+\beta(m+1)q^T\partial^{-1}p^T\\
I_m+\beta(m+1)p\partial^{-1}q&-\beta(m+1)p\partial^{-1}p^T
\end{pmatrix}.
\end{align}
Among the soliton hierarchy \eqref{SH1}, the first nontrivial nonlinear systems is
\begin{equation}\left\{\begin{array}{l}
p_{t_2}=\frac{1}{m+1}(p_{xx}-2rp)-\beta[2(pq)_xp+(m+1)\beta r^2p],\\
q_{t_2}=-\frac{1}{m+1}(q_{xx}-2rq)-\beta[2q(pq)_x-(m+1)\beta r^2q].\end{array}\right.
\end{equation}

\subsection{Bi-Hamiltonian structures and Liouville integrability}
In order to establish bi-Hamiltonian structures of the generalized multi-component AKNS hierarchy \eqref{SH1}, we shall use the trace identity
\begin{align}
\frac{\delta}{\delta u}\int \mbox{tr}\left(W\frac{\partial U}{\partial \lambda}\right)dx=\lambda^{-\gamma}\frac{\partial}{\partial\lambda}\left[\lambda^\gamma \mbox{tr}\left(W\frac{\partial U}{\partial u}\right)\right].
\end{align}
It is direct to calculate that
\begin{equation}\label{TI1}\left\{\begin{array}{l}
\mbox{tr}\left(W\frac{\partial U}{\partial \lambda}\right)=-ma+\mbox{tr}(d),\\
\mbox{tr}\left(W\frac{\partial U}{\partial u}\right)
=-\beta m a\begin{pmatrix}p\\ q^T\end{pmatrix}+
\begin{pmatrix}c\\ b^T\end{pmatrix}+\beta\mbox{tr}(d)\begin{pmatrix}p\\q^T\end{pmatrix}.
\end{array}\right.
\end{equation}
By substituting \eqref{TI1} into the trace identity and balancing the coefficients of each power of $\lambda$, we have
%\begin{align}
%&\frac{\partial U}{\partial \lambda}=\begin{pmatrix}-m&0\\0&I_m\end{pmatrix},\, \mbox{tr}\left(W\frac{\partial U}{\partial \lambda}\right)=-ma+\mbox{tr}(d)=\sum\limits_{k\ge 0}\left(-ma^{(k)}+\sum\limits_{i=1}^md_{ii}^{(k)}\right)\lambda^{-k},\\
%&\mbox{tr}\left(W\frac{\partial U}{\partial u}\right)
%=-\beta m a\begin{pmatrix}p\\ q^T\end{pmatrix}+
%\begin{pmatrix}c\\ b^T\end{pmatrix}+\beta\mbox{tr}(d)\begin{pmatrix}p\\q^T\end{pmatrix}
%=\sum\limits_{k\ge 0}\left(-\beta m\begin{pmatrix}p\\ q^T\end{pmatrix}a^{(k)}+\begin{pmatrix}c^{(k)}\\ {b^{(k)}}^T\end{pmatrix}+\beta\begin{pmatrix}p\\q^T\end{pmatrix}\sum\limits_{i=1}^md_{ii}^{(k)} \right)\lambda^{-k}
%\end{align}
\begin{align*}
\frac{\delta}{\delta u}\int\left(-ma^{(k+1)}+\sum\limits_{i=1}^md_{ii}^{(k+1)}\right)dx=(\gamma-k)\left(-\beta m\begin{pmatrix}p\\ q^T\end{pmatrix}a^{(k)}+\begin{pmatrix}c^{(k)}\\ {b^{(k)}}^T\end{pmatrix}+\beta\begin{pmatrix}p\\q^T\end{pmatrix}\sum\limits_{i=1}^md_{ii}^{(k)} \right), k\ge 0.
\end{align*}
%\begin{align}
%&\mbox{tr}\left(W\frac{\partial U}{\partial \lambda}\right)=-ma+\mbox{tr}(d)=-ma+\partial^{-1}(bp-qc)
%=\sum\limits_{k\ge 0}[-ma^{(k)}+\partial^{-1}(b^{(k)}p-qc^{(k)})]\lambda^{-k},\\
%&\mbox{tr}\left(W\frac{\partial U}{\partial u}\right)
%=-\beta m a\begin{pmatrix}p\\ q^T\end{pmatrix}+
%\begin{pmatrix}c\\ b^T\end{pmatrix}+\beta\mbox{tr}(d)\begin{pmatrix}p\\q^T\end{pmatrix}\\
%&=\sum\limits_{k\ge 0}\left(-\beta m\begin{pmatrix}p\\ q^T\end{pmatrix}a^{(k)}+\begin{pmatrix}c^{(k)}\\ {b^{(k)}}^T\end{pmatrix}+\beta\begin{pmatrix}p\\q^T\end{pmatrix}\partial^{-1}(b^{(k)}p-qc^{(k)}) \right)\lambda^{-k}
%\end{align}
To determine the constant $\gamma$, we can simply let $k=1$ in the above equation and get $\gamma=0$. Thus we obtain
\begin{align}
\frac{\delta}{\delta u}\mathcal{H}_k=\left(-\beta m\begin{pmatrix}p\\ q^T\end{pmatrix}a^{(k)}+\begin{pmatrix}c^{(k)}\\ {b^{(k)}}^T\end{pmatrix}+\beta\begin{pmatrix}p\\q^T\end{pmatrix}\sum\limits_{i=1}^md_{ii}^{(k)} \right),\quad m\ge 0
\end{align}
with the Hamiltonian functionals being defined by
%\begin{align}
%-\frac{\delta}{\delta u}\int\frac{\left(-ma^{(k+1)}+\sum\limits_{i=1}^md_{ii}^{(k+1)}\right)}{k}dx=\left(-\beta m\begin{pmatrix}p\\ q^T\end{pmatrix}a^{(k)}+\begin{pmatrix}c^{(k)}\\ {b^{(k)}}^T\end{pmatrix}+\beta\begin{pmatrix}p\\q^T\end{pmatrix}\sum\limits_{i=1}^md_{ii}^{(k)} \right)=G^{(k)}, k\ge 1
%\end{align}
\begin{align}
\mathcal{H}_0=\beta m(m+1)\int r dx,\quad\mathcal{H}_k=\frac{\delta}{\delta u}\int\frac{\left(ma^{(k+1)}-\sum\limits_{i=1}^md_{ii}^{(k+1)}\right)}{k}dx,\quad m\ge 1.
\end{align}

Denote
\begin{align}
G^{(k)}=\left(-\beta m\begin{pmatrix}p\\ q^T\end{pmatrix}a^{(k)}+\begin{pmatrix}c^{(k)}\\ {b^{(k)}}^T\end{pmatrix}+\beta\begin{pmatrix}p\\q^T\end{pmatrix}\sum\limits_{i=1}^md_{ii}^{(k)} \right),
\end{align}
then we have
\begin{align}
G^{(k)}=-\beta m\begin{pmatrix}p\\ q^T\end{pmatrix}a^{(k)}+\begin{pmatrix}c^{(k)}\\ {b^{(k)}}^T\end{pmatrix}+\beta\begin{pmatrix}p\\ q^T\end{pmatrix}\partial^{-1}(b^{(k)}p-qc^{(k)})=N^{-1}\begin{pmatrix}
c^{(k)}\\{b^{(k)}}^T
\end{pmatrix}
\end{align}
with
\begin{align}
N^{-1}=\begin{pmatrix}
I_m-\beta(m+1)p\partial^{-1}q&\beta(m+1)p\partial^{-1}p^T\\
-\beta(m+1)q^T\partial^{-1}q&I_m+\beta(m+1)q^T\partial^{-1}p^T
\end{pmatrix}.
\end{align}
Conversely, we have
\begin{align}
\begin{pmatrix}
c^{(k)}\\{b^{(k)}}^T
\end{pmatrix}=NG^{(k)}, \quad N=\begin{pmatrix}
I_m+\beta(m+1)p\partial^{-1}q&-\beta(m+1)p\partial^{-1}p^T\\
\beta(m+1)q^T\partial^{-1}q&I_m-\beta(m+1)q^T\partial^{-1}p^T
\end{pmatrix}.
\end{align}

%
%\begin{align}
%G^{(k)}&=-\beta m\begin{pmatrix}p\\ q^T\end{pmatrix}a^{(k)}+\begin{pmatrix}c^{(k)}\\ {b^{(k)}}^T\end{pmatrix}+\beta\begin{pmatrix}p\\ q^T\end{pmatrix}\partial^{-1}(b^{(k)}p-qc^{(k)})\\
%&=-\beta m\begin{pmatrix}p\\ q^T\end{pmatrix}\partial^{-1}(qc^{(k)}-b^{(k)}p)+\begin{pmatrix}c^{(k)}\\ {b^{(k)}}^T\end{pmatrix}+\beta\begin{pmatrix}p\\q^T\end{pmatrix}\partial^{-1}(b^{(k)}p-qc^{(k)})\\
%&=\begin{pmatrix}
%-\beta(m+1)p\partial^{-1}(qc^{(k)})+\beta(m+1)p\partial^{-1}(b^{(k)}p)+c^{(k)}\\
%-\beta(m+1)q^T \partial^{-1}(qc^{(k)})+\beta(m+1)q^T\partial^{-1}(b^{(k)}p)+{b^{(k)}}^T
%\end{pmatrix}\\
%&=\begin{pmatrix}
%I_m-\beta(m+1)p\partial^{-1}q&\beta(m+1)p\partial^{-1}p^T\\
%-\beta(m+1)q^T\partial^{-1}q&I_m+\beta(m+1)q^T\partial^{-1}p^T
%\end{pmatrix}
%\begin{pmatrix}
%c^{(k)}\\{b^{(k)}}^T
%\end{pmatrix}=N^{-1}\begin{pmatrix}
%c^{(k)}\\{b^{(k)}}^T
%\end{pmatrix}\\
%&N^{-\dagger}=\begin{pmatrix}
%I_m+\beta(m+1)q^T\partial^{-1}p^T&\beta(m+1)q^T\partial^{-1}q\\
%-\beta(m+1)p\partial^{-1}p^T&I_m-\beta(m+1)p\partial^{-1}q
%\end{pmatrix},\\
%\begin{pmatrix}
%c^{(k)}\\{b^{(k)}}^T
%\end{pmatrix}&
%=\begin{pmatrix}
%I_m+\beta(m+1)p\partial^{-1}q&-\beta(m+1)p\partial^{-1}p^T\\
%\beta(m+1)q^T\partial^{-1}q&I_m-\beta(m+1)q^T\partial^{-1}p^T
%\end{pmatrix}
%G^{(k)}=NG^{(k)},\\
%&N^{\dagger}=\begin{pmatrix}
%I_m-\beta(m+1)q^T\partial^{-1}p^T&-\beta(m+1)q^T\partial^{-1}q\\
%\beta(m+1)p\partial^{-1}p^T&I_m+\beta(m+1)p\partial^{-1}q
%\end{pmatrix}
%\end{align}

It follows from the above results that the soliton hierarchy \eqref{SH1} has the Hamiltonian structures
\begin{align}
u_{t_n}=K_n=J\frac{\delta \mathcal{H}_{n+1}}{\delta u},
\end{align}
where $J$ is a Hamiltonian operator given by
\begin{align}
J=RN=(m+1)\begin{pmatrix}-2\beta(m+1)q^T\partial^{-1}q&-I_m+2\beta(m+1)q^T\partial^{-1}p^T\\
I_m+2\beta(m+1)p\partial^{-1}q&-2\beta(m+1)p\partial^{-1}p^T
\end{pmatrix}.
\end{align}

%\subsection{non-isospectral problem}
It is obvious that $G^{(n+1)}=\Psi G^{(n)}$ with $\Psi=N^{-1}LN$.
From $K_{n+1}=\Phi K_n,\,\, n\ge 0$, and $J\Psi=\Phi J$, we derive a common recursion operator \cite{PO} for the generalized multicomponent AKNS hierarchy \eqref{SH1} given by
\begin{align}
\Phi=\Psi^{\dagger}=N^\dagger L^\dagger N^{-\dagger},
\end{align}
where $\Psi^\dagger$ denotes the adjoint operator of $\Psi$. The explicit form of $\Psi$ is given by
\begin{equation*}\left\{\begin{array}{l}
\Phi_{11}=\beta^2(m+1)q^T\partial^{-1}q\partial p\partial^{-1}p^T+\beta q^T\partial^{-1}p^T\{[\partial+\beta(m+1)r]I_m+\beta(m+1)\partial q^T\partial^{-1}p^T\}\\
\quad\qquad+\frac{1}{m+1}\left\{q^T\partial^{-1}p^T-\left[\partial+\beta(m+1)r-\sum\limits_{i=1}^{m}q_i\partial^{-1}p_i\right]I_m-\beta(m+1)[\partial+\beta(m+1)r]q^T\partial^{-1}p^T\right\},\\
\Phi_{12}=\frac{1}{m+1}\{q^T\partial^{-1}q+(q^T\partial^{-1}q)^T-\beta(m+1)[\partial+\beta(m+1)r]q^T\partial^{-1}q\}\\
\quad\qquad+\beta^2(m+1)q^T\partial^{-1}p^T\partial q^T\partial^{-1}q-\beta q^T\partial^{-1}q\{[\partial-\beta(m+1)r]I_m-\beta(m+1)\partial p\partial^{-1}q\},\\
\Phi_{21}=-\frac{1}{m+1}\{p\partial^{-1}p^T+(p\partial^{-1}p^T)^T+\beta(m+1)[\partial-\beta(m+1)r]p\partial^{-1}p^T\}\\
\quad\qquad-\beta^2(m+1)p\partial^{-1}q\partial p\partial^{-1}p^T-\beta p\partial^{-1}p^T\{[\partial+\beta(m+1)r]I_m+\beta(m+1)\partial q^T\partial^{-1}p^T\},\\
\Phi_{22}=-\beta^2(m+1)p\partial^{-1}p^T\partial q^T\partial^{-1}q+\beta p\partial^{-1}q\{[\partial-\beta(m+1)r]I_m-\beta(m+1)\partial p\partial^{-1}q\}\\
\quad\qquad+\frac{1}{m+1}\left\{-p\partial^{-1}q+\left[\partial-\beta(m+1)r-\sum\limits_{i=1}^mp_i\partial^{-1}q_i\right]I_m-\beta(m+1)[\partial-\beta(m+1)r]p\partial^{-1}q\right\}.
\end{array}\right.\
\end{equation*}

It is now a straightforward computation to show that all members in the soliton hierarchy \eqref{SH1} are bi-Hamiltonian:
\begin{align}\label{BH1}
u_{t_n}=K_n=J\frac{\delta \mathcal{H}_{n+1}}{\delta u}=M\frac{\delta \mathcal{H}_{n}}{\delta u},\quad m\ge 0,
\end{align}
where the second Hamiltonian operator $M$ given by $M=\Phi J$. The entries of $M$ are defined by
\begin{equation*}\left\{\begin{array}{l}
M_{11}=\beta(m+1)\left[(\partial +\beta(m+1)r)I_m-\beta(m+1)q^T\partial^{-1}p^T\partial\right]q^T\partial^{-1}q\\
\qquad\quad+q^T\partial^{-1}q+(q^T\partial^{-1}q)^T-\beta(m+1)q^T\partial^{-1}q[(\partial-\beta(m+1)r)I_m+\beta(m+1)\partial p\partial^{-1}q],\\
M_{12}=-q^T\partial^{-1}p^T+\left[\partial+\beta(m+1)r-\sum\limits_{i=1}^mq_i\partial^{-1}p_i\right]I_m-\beta(m+1)[\partial+\beta(m+1)r]q^T\partial^{-1}p^T\\
\qquad\quad+\beta^2(m+1)^2q^T\partial^{-1}q\partial p\partial^{-1}p^T-\beta(m+1)q^T\partial^{-1}p^T[(\partial+\beta(m+1)r)I_m+\beta(m+1)\partial q^T\partial^{-1}p^T],\\
M_{21}=-p\partial^{-1}q+\left[\partial-\beta(m+1)r-\sum\limits_{i=1}^mp_i\partial^{-1}q_i\right]I_m+\beta(m+1)[\partial-\beta(m+1)r]p\partial^{-1}q\\
\qquad\quad+\beta^2(m+1)^2p\partial^{-1}p^T\partial q^T\partial^{-1}q+\beta(m+1)p\partial^{-1}q[(\partial-\beta(m+1)r)I_m+\beta(m+1)\partial p\partial^{-1}q],\\
M_{22}=-\beta(m+1)\left[(\partial -\beta(m+1)r)I_m+\beta(m+1)p\partial^{-1}q\partial\right]p\partial^{-1}p^T\\
\qquad\quad+p\partial^{-1}p^T+(p\partial^{-1}p^T)^T+\beta(m+1)p\partial^{-1}p^T[(\partial+\beta(m+1)r)I_m-\beta(m+1)\partial q^T\partial^{-1}p^T].
\end{array}\right.
\end{equation*}

The soliton hierarchy \eqref{SH1} is Liouville integrable, upon noticing that the vector fields $K_n$,\quad $ n\ge 1$ possess distinct differential orders and that the common conserved functionals $\{\mathcal{H}_n\}_{n=0}^{\infty}$ and symmetries $\{\mathcal{K}_n\}_{n=0}^{\infty}$ commute:
\begin{equation}\left\{\begin{array}{l}
\{\mathcal{H}_l,\mathcal{H}_m\}_{J}=\int ( \frac{\delta\mathcal{H}_l}{\delta u})^T J\frac{\delta \mathcal{H}_m}{\delta u}dx=0,\\
\{\mathcal{H}_l,\mathcal{H}_m\}_{M}=\int ( \frac{\delta\mathcal{H}_l}{\delta u})^T M\frac{\delta \mathcal{H}_m}{\delta u}dx=0,\end{array}\right. \, m\ge 0,
\end{equation}
and
\begin{align}
[K_l,K_m]=K_l'(u)[K_m]-K_m'(u)[K_l]=0,\ l,m\ge 0.
\end{align}

\begin{remark}
When we set $m=1$ and $\beta=0$ throughout Section 2, we can recover a series of results on the classical AKNS spectral problem considered in \cite{AKNS}.
\end{remark}
\begin{remark}
By setting $m=1$ and $\beta=\mu$ in Section 2, we can obtain the results on the matrix spectral problem proposed by Yan etc. in \cite{YZY}.
\end{remark}
\begin{remark}
When $\beta=0$, the generalized multicomponent AKNS spectral problem \eqref{GAKNS1}, its corresponding soliton hierarchy \eqref{SH1} and bi-Hamiltonian structures \eqref{BH1} are reduced to the ones considered in \cite{MA6}.
\end{remark}

%The next part is about so(n,r)

\section{A hierarchy of generalized multicomponent AKNS-type equations associated with antisymmetric matrices of $(m+2)$-th order}
\subsection{A hierarchy of generalized multicomponent AKNS-type equations}
In this section, we will consider the following matrix spectral problem which we call the generalized multicomponent AKNS-type spectral problem
\begin{align}\label{GAKNS2}
\phi_x=U(u,\lambda)\phi=\begin{pmatrix}
0&q&m(\lambda+\beta r)\\
-q^T&0&-p\\
-m(\lambda+\beta r)&p^T&0
\end{pmatrix}\phi,
\end{align}
where $m$ is an arbitrary natural number, $\lambda$ is a spectral parameter, $I_m$ is an $m\times m$ identity matrix and
\begin{align}
q=(q_1,q_2,\cdots,q_m),\ p=(p_1,p_2,\cdots,p_m)^T,\ r=qq^T+p^Tp,\ u=(p^T,\ q)^T.
\end{align}
%\begin{align}
%W&=\begin{pmatrix}0&c&a\\
%-c^T&d&-b\\
%-a&b^T&0\end{pmatrix}, r=p^Tp+qq^T,u=\begin{pmatrix}p\\q^T\end{pmatrix}
%\end{align}

Following the standard procedure, let us first solve the stationery zero-curvature equation
\begin{align}
W_x=[U,W],\quad W=\begin{pmatrix}0&c&a\\
-c^T&d&-b\\
-a&b^T&0\end{pmatrix},
\end{align}
which gives
%\begin{align}
%&qc^T=cq^T,b^Tp=p^Tb,(pb^T)^T=bp^T,(q^Tc)^T=c^Tq,\\
%&(qb)^T=qb=b^Tq^T,(cp)^T=cp=p^Tc^T,d^T=-d\\
%&c_x=qd-ap^T+m(\lambda+\beta r)b^T,\\
%&a_x=-qb+cp,\\
%&-{c^T}_x=pa+dq^T-m(\lambda+\beta r)b,\\
%&d_x=-q^Tc-pb^T+c^Tq+bp^T,\\
%&-b_x=-q^Ta+m(\lambda+\beta r)c^T+dp,\\
%&-a_x=-p^Tc^T+b^Tq^T,\\
%&{b^T}_x=-m(\lambda+\beta r)c+p^Td+aq,\\
%&{d_x}^T=-c^Tq-bp^T+q^Tc+pb^T=-d_x.
%\end{align}
\begin{equation}\label{SZE2}\left\{\begin{array}{l}
a_x=-qb+cp,\\
b_x=q^Ta-m(\lambda+\beta r)c^T-dp,\\
c_x=qd-ap^T+m(\lambda+\beta r)b^T,\\
d_x=-q^Tc-pb^T+c^Tq+bp^T.\end{array} \right.
\end{equation}
A direct calculation also tells us that
\begin{align}
{d_x}^T=-c^Tq-bp^T+q^Tc+pb^T=-d_x.
\end{align}
Upon letting
\begin{align}
W=\sum\limits_{k\ge 0}W_k\lambda^{-k}=\sum\limits_{k\ge 0}\begin{pmatrix}0&c^{(k)}&a^{(k)}\\
-{c^{(k)}}^T&d^{(k)}&-b^{(k)}\\
-a^{(k)}&{b^{(k)}}^T&0\end{pmatrix}\lambda^{-k},
\end{align}
with $b^{(k)},c^{(k)}$ and $d^{(k)}$ being assumed to be
\begin{align}
b^{(k)}=(b_1^{(k)},b_2^{(k)},\cdots,b_m^{(k)})^T,\quad c^{(k)}=(c_1^{(k)},c_2^{(k)},\cdots,c_m^{(k)}),\quad d^{(k)}=(d_{ij}^{(k)})_{m\times m},
\end{align}
the system \eqref{SZE2} leads to
\begin{equation}\left\{\begin{array}{l}
c^{(0)}=0,\,b^{(0)}=0,\, a^{(k)}_x=c^{(k)}p-qb^{(k)},\\
d^{(k)}_x={c^{(k)}}^Tq+b^{(k)}p^T-q^Tc^{(k)}-p{b^{(k)}}^T,\\
{c^{(k+1)}}^T=\frac{1}{m}(q^Ta^{(k)}-m\beta r{c^{(k)}}^T-d^{(k)}p-b^{(k)}_x),\\
{b^{(k+1)}}^T=\frac{1}{m}(c^{(k)}_x-qd^{(k)}+a^{(k)}p^T-m\beta r{b^{(k)}}^T),\end{array}\right.
\end{equation}
which yields the recursion relation
%\begin{align}
%&a^{(k)}_x=c^{(k)}p-qb^{(k)},\\
%&{c^{(k+1)}}^T=\frac{1}{m}(q^Ta^{(k)}-m\beta r{c^{(k)}}^T-d^{(k)}p-b^{(k)}_x),\\
%&{b^{(k+1)}}^T=\frac{1}{m}(c^{(k)}_x-qd^{(k)}+a^{(k)}p^T-m\beta r{b^{(k)}}^T),\\
%&d^{(k)}_x={c^{(k)}}^Tq+b^{(k)}p^T-q^Tc^{(k)}-p{b^{(k)}}^T,\\
%&(p^Tp)_x=2p_1p_{1x}+2p_2p_{2x}+\cdots=2p^Tp_x=2p^T_xp,\\
%&(p^Tp)_x=p^T_xp+p^Tp_x=2(p^Tp_x)=2(p^T_xp),p^T_xp=p^T_xp
%\end{align}
%The recursion relation is
\begin{align}
\begin{pmatrix}b^{(k+1)}\\{c^{(k+1)}}^T\end{pmatrix}=L\begin{pmatrix}b^{(k)}\\{c^{(k)}}^T\end{pmatrix}
\end{align}
with
\begin{align*}
L=\frac{1}{m}\begin{pmatrix}
(\sum\limits_{i=1}^mq_i\partial^{-1}p_i-m\beta r)I_m-p\partial^{-1}q-(q^T\partial^{-1}p^T)^T&(\partial+\sum\limits_{i=1}^mq_i\partial^{-1}q_i)I_m+p\partial^{-1}p^T-(q^T\partial^{-1}q)^T\\
-(\partial+\sum\limits_{i=1}^mp_i\partial^{-1}p_i)I_m-q^T\partial^{-1}q+(p\partial^{-1}p^T)^T&-(m\beta r+\sum\limits_{i=1}^mp_i\partial^{-1}q_i)I_m+q^T\partial^{-1}p^T+(p\partial^{-1}q)^T
\end{pmatrix}.
\end{align*}
%
%\begin{align}
%L^{\dagger}==\frac{1}{m}\begin{pmatrix}
%-(\sum\limits_{i=1}^mp_i\partial^{-1}q_i+m\beta r)I_m+q^T\partial^{-1}p^T+(p\partial^{-1}q)^T&(\partial+\sum\limits_{i=1}^mp_i\partial^{-1}p_i)I_m+q^T\partial^{-1}q-(p\partial^{-1}p^T)^T\\
%-(\partial+\sum\limits_{i=1}^mq_i\partial^{-1}q_i)I_m-p\partial^{-1}p^T+(q^T\partial^{-1}q)^T&-(m\beta r-\sum\limits_{i=1}^mq_i\partial^{-1}p_i)I_m-p\partial^{-1}q-(q^T\partial^{-1}p^T)^T
%\end{pmatrix}
%\end{align}
%

By choosing the initial values to be
\begin{align}
a^{(0)}=-m,\quad d^{(0)}=0,
\end{align}
and requiring that
\begin{align}
W_k|_{u=0}=0,\quad k\ge 1,
\end{align}
we can determine the sequence $\{a^{(k)},b^{(k)},c^{(k)},d^{(k)}|\,\,k\ge 1\}$ uniquely. The first few sets can be computed as
\begin{equation*}\left\{\begin{array}{l}
a^{(1)}=0, b^{(1)}=-p,c^{(1)}=-q,d^{(1)}=0,\\
a^{(2)}=\frac{1}{2m}(qq^T+p^Tp),\quad b^{(2)}=\beta rp-\frac{q^T_x}{m},\\
c^{(2)}=\beta rq+\frac{p^T_x}{m},\quad d^{(2)}=\frac{1}{m}(pq-q^Tp^T),\\
a^{(3)}=\frac{1}{m^2}(q_xp-qp_x)-\frac{1}{m}\beta r^2, \\
d^{(3)}=\frac{1}{m^2}(q_x^Tq-q^Tq_x+p_xp^T-pp_x^T)-\frac{2}{m}\beta r(pq-q^Tp^T),\\
b^{(3)}=\frac{1}{m}\left(\frac{3}{2m}pqq^T+\frac{1}{2m}pp^Tp-\frac{1}{m}q^Tqp+\frac{1}{m}p_{xx}-m\beta^2 r^2p+2\beta rq_x^T+\beta r_xq^T\right),\\
c^{(3)}=\frac{1}{m}\left(\frac{3}{2m}p^Tpq+\frac{1}{2m}qq^Tq-\frac{1}{m}qpp^T+\frac{1}{m}q_{xx}-m\beta^2r^2q-2\beta rp_x^T-\beta r_xp^T\right).\end{array}\right.
\end{equation*}

To construct the auxiliary spectral problem, let us take
\begin{align}\label{AS2}
V^{(n)}=(\lambda^nW)_{+}+\Delta_n=\sum\limits_{k\ge 0}^{n}W_k\lambda^{n-k}+\begin{pmatrix}
0&0&f_n\\0&0&0\\-f_n&0&0\end{pmatrix},\quad n\ge 0
\end{align}
with $f_n=-2m^2\beta a^{(n+1)}$. The compatibility condition of \eqref{GAKNS2} and \eqref{AS2} generates the zero-curvature equation
\begin{align}
U_{t_n}-V^{(n)}_x+[U,V^{(n)}]=0,
\end{align}
which is equivalent to
\begin{equation}\left\{\begin{array}{l}
f_{nx}=m\beta r_{t_n},\\%=-2m^2\beta a^{(n+1)}_x,f_n=-2m^2\beta a^{(n+1)},r_{t_n}=-2ma_x^{(n+1)},\\
%&q_{t_n}=f_np^T+m{b^{(n+1)}}^T,\\%=c_x^{(n)}+f_np^T-qd^{(n)}-m\beta r{b^{(n)}}^T+a^{(n)}p^T
%%&q_{t_n}=-2m^2\beta \partial^{-1}(c^{(n+1)}p)p^T+2m^2\beta\partial^{-1}(qb^{(n+1)})p^T+m{b^{(n+1)}}^T,\\
%&q_{t_n}^T=-2m^2\beta p\partial^{-1}(p^T{c^{(n+1)}}^T)+2m^2\beta p\partial^{-1}({b^{(n+1)}}^Tq^T)+mb^{(n+1)},\\
%&p_{t_n}=b_x^{(n)}-q^Tf_n-q^Ta^{(n)}+d^{(n)}p+m\beta r{c^{(n)}}^T=-q^Tf_n-m{c^{(n+1)}}^T,\\
%&p_{t_n}=2m^2\beta q^T\partial^{-1}(c^{(n+1)}p-qb^{(n+1)})-m{c^{(n+1)}}^T,\\
q_{t_n}=f_np^T+m{b^{(n+1)}}^T,\\
p_{t_n}=-q^Tf_n-m{c^{(n+1)}}^T.\end{array}\right.
\end{equation}
Actually we can solve $f_n$ to get $f_n=-2m^2\beta a^{(n+1)}$. Finally we derive the generalized multicomponent AKNS-type hierarchy
\begin{align}\label{SH2}
\begin{pmatrix}p\\q^T\end{pmatrix}_{t_n}=R\begin{pmatrix}b^{(n+1)}\\{c^{(n+1)}}^T\end{pmatrix}=m
\begin{pmatrix}
-2m\beta q^T\partial^{-1}q&-I_m+2m\beta q^T\partial^{-1}p^T\\
I_m+2m\beta p\partial^{-1}q&-2m\beta p\partial^{-1}p^T
\end{pmatrix}
\begin{pmatrix}b^{(n+1)}\\{c^{(n+1)}}^T\end{pmatrix}.
\end{align}
The first nontrivial nonlinear system in \eqref{SH2} is
\begin{equation*}\left\{\begin{array}{l}
p_{t_n}=-m\beta^2 r^2q^T-\frac{1}{2m}(3q^Tp^Tp+q^Tqq^T-2pp^Tq^T)+\beta(2q_xpq^T-2qp_xq^T+2rp_x+r_xp)-\frac{1}{m}q_{xx}^T,\\
q_{t_n}=m\beta^2r^2p^T+\frac{1}{2m}(3qq^Tp^T+p^Tpp^T-2p^Tq^Tq)+\beta(2qp_xp^T-2q_xpp^T+2rq_x+r_xq)+\frac{1}{m}p_{xx}^T. \end{array}\right.
\end{equation*}

\subsection{Bi-Hamiltonian structures and Liouville integrability}
To construct the Hamiltonian structure of the generalized multicomponent AKNS-type hierarchy \eqref{SH2}, we shall use the trace identity
\begin{align}\label{TI}
\frac{\delta}{\delta u}\int \mbox{tr}\left(W\frac{\partial U}{\partial \lambda}\right)dx=\lambda^{-\gamma}\frac{\partial}{\partial\lambda}\left[\lambda^\gamma \mbox{tr}\left(W\frac{\partial U}{\partial u}\right)\right].
\end{align}
By detailed calculations, we have
\begin{align}\label{TI3}
\mbox{tr}\left(W\frac{\partial U}{\partial \lambda}\right)=-2ma,\quad
\mbox{tr}\left(W\frac{\partial U}{\partial u}\right)
=-4m\beta\begin{pmatrix}p\\q^T\end{pmatrix} a-2\begin{pmatrix}b\\c^T\end{pmatrix}.
\end{align}
Substituting \eqref{TI3} into the trace identity \eqref{TI} and balancing coefficients of each power of $\lambda$, we have
%\begin{align}
%&\frac{\partial U}{\partial \lambda}=\begin{pmatrix}0&0&m\\0&0&0\\-m&0&0\end{pmatrix},\, \mbox{tr}\left(W\frac{\partial U}{\partial \lambda}\right)=-2ma,\\
%&\mbox{tr}\left(W\frac{\partial U}{\partial u}\right)
%=-4m\beta\begin{pmatrix}p\\q^T\end{pmatrix} a-2\begin{pmatrix}b\\c^T\end{pmatrix}
%\end{align}
\begin{align}
\frac{\delta}{\delta u}\int ma^{(k+1)}dx=(\gamma-k)\left(2m\beta\begin{pmatrix}p\\q^T\end{pmatrix}a^{(k)}+
\begin{pmatrix}b^{(k)}\\{c^{(k)}}^T\end{pmatrix}\right),\quad k\ge 0.
\end{align}
To determine the constant $\gamma$, we can simply set $k=1$ to get $\gamma=0$. In this way we have
\begin{align}
\frac{\delta}{\delta u}\mathcal{H}_k=\left(2m\beta\begin{pmatrix}p\\q^T\end{pmatrix}a^{(k)}+
\begin{pmatrix}b^{(k)}\\{c^{(k)}}^T\end{pmatrix}\right),\quad k\ge 0
\end{align}
with
\begin{align}
\mathcal{H}_0=-\beta m^2\int (p^Tp+qq^T) dx,\quad\mathcal{H}_k=-\int \frac{m}{k}a^{(k+1)}dx, \quad k\ge 1.
\end{align}

Denote that
\begin{align}G^{(k)}=\left(2m\beta\begin{pmatrix}p\\q^T\end{pmatrix}a^{(k)}+
\begin{pmatrix}b^{(k)}\\{c^{(k)}}^T\end{pmatrix}\right)=\begin{pmatrix}
2m\beta pa^{(k)}+b^{(k)}\\
2m\beta q^Ta^{(k)}+{c^{(k)}}^T
\end{pmatrix},
\end{align}
then we have
\begin{align}
G^{(k)}&=N^{-1}\begin{pmatrix}
b^{(k)}\\{c^{(k)}}^T\end{pmatrix},\quad N^{-1}=\begin{pmatrix}
I_m-2m\beta p\partial^{-1}q&2m\beta p\partial^{-1}p^T\\
-2m\beta q^T\partial^{-1}q&I_m+2m\beta q^T\partial^{-1}p^T
\end{pmatrix}.
\end{align}
%
%\begin{align}
%N^{-\dagger}=\begin{pmatrix}
%I_m+2m\beta q^T\partial^{-1}p^T&2m\beta q^T\partial^{-1}q\\
%-2m\beta p\partial^{-1}p^T&I_m-2m\beta p\partial^{-1}q
%\end{pmatrix}
%\end{align}
Conversely, we have
\begin{align}
\begin{pmatrix}
b^{(k)}\\{c^{(k)}}^T
\end{pmatrix}&=NG^{(k)}, \quad N=\begin{pmatrix}
I_m+2m\beta p\partial^{-1}q&-2m\beta p\partial^{-1}p^T\\
2m\beta q^T\partial^{-1}q&I_m-2m\beta q^T\partial^{-1}p^T
\end{pmatrix}.
%
%\begin{pmatrix}
%2m\beta p\partial^{-1}(c^{(k)}p-qb^{(k)})+b^{(k)}\\
%2m\beta q^T\partial^{-1}(c^{(k)}p-qb^{(k)})+{c^{(k)}}^T
%\end{pmatrix}\\
%&=\begin{pmatrix}
%I_m+2m\beta p\partial^{-1}q&-2m\beta p\partial^{-1}p^T\\
%2m\beta q^T\partial^{-1}q&I_m-2m\beta q^T\partial^{-1}p^T
%\end{pmatrix}
%\begin{pmatrix}
%2m\beta p\partial^{-1}(c^{(k)}p-qb^{(k)})+b^{(k)}\\
%2m\beta q^T\partial^{-1}(c^{(k)}p-qb^{(k)})+{c^{(k)}}^T
%\end{pmatrix}
\end{align}
%
%\begin{align}
%N^\dagger=\begin{pmatrix}
%I_m-2m\beta q^T\partial^{-1}p^T&-2m\beta q^T\partial^{-1}q\\
%2m\beta p\partial^{-1}p^T&I_m+2m\beta p\partial^{-1}q
%\end{pmatrix}
%\end{align}
%
%

From the obtained results, it follows that the soliton hierarchy \eqref{SH2} has Hamiltonian structures
\begin{align}
u_{t_n}=K_n=J\frac{\delta \mathcal{H}_{n+1}}{\delta u},%=RNG^{(n+1)}=JG^{(n+1)}
\end{align}
where $J=RN$ is a Hamiltonian operator defined by
\begin{align}
J=m\begin{pmatrix}-4m\beta q^T\partial^{-1}q&-I_m+4m\beta q^T\partial^{-1}p^T\\
I_m+4\beta m p\partial^{-1}q&-4\beta m p\partial^{-1}p^T
\end{pmatrix}
\end{align}

%\subsection{non-isospectral problem}
Similar to Section 2, we can derive a common recursion operator $\Psi=N^\dagger L^\dagger N^{-\dagger}$ for the generalized multicomponent AKNS-type hierarchy \eqref{SH2} given by
\begin{equation*}\left\{\begin{array}{l}
\Phi_{11}=2m\beta^2 q^T\partial^{-1}p^T(r I_m+2\partial p\partial^{-1}p^T)+2\beta q^T\partial^{-1}q(\partial I_m+2m\beta \partial q^T\partial^{-1}p^T)\\
\qquad\quad+\frac{1}{m}\left(q^T\partial^{-1}p^T+(p\partial^{-1}q)^T-(m\beta r+\sum\limits_{i=1}^{m}p_i\partial^{-1}q_i)I_m-2m\beta(\partial p\partial^{-1}p^T+m\beta rq^T\partial^{-1}p^T)\right),\\
\Phi_{12}=2m\beta^2 q^T\partial^{-1}q(r I_m+2\partial q^T\partial^{-1}q)-2\beta q^T\partial^{-1}p^T(\partial I_m-2m\beta \partial p\partial^{-1}q)\\
\qquad\quad+\frac{1}{m}\left(q^T\partial^{-1}q-(p\partial^{-1}p^T)^T+(\partial+\sum\limits_{i=1}^{m}p_i\partial^{-1}p_i)I_m-2m\beta(\partial p\partial^{-1}q+m\beta rq^T\partial^{-1}q)\right),\\
\Phi_{21}=-2m\beta^2 p\partial^{-1}p^T(r I_m+2\partial p\partial^{-1}p^T)-2\beta p\partial^{-1}q(\partial I_m+2m\beta \partial q^T\partial^{-1}p^T)\\
\qquad\quad+\frac{1}{m}\left((q^T\partial^{-1}q)^T-p\partial^{-1}p^T-(\partial+\sum\limits_{i=1}^{m}q_i\partial^{-1}q_i)I_m-2m\beta(\partial q^T\partial^{-1}p^T-m\beta rp\partial^{-1}p^T)\right),\\
\Phi_{22}=-2m\beta^2 p\partial^{-1}q(r I_m+2\partial q^T\partial^{-1}q)+2\beta p\partial^{-1}p^T(\partial I_m-2m\beta \partial p\partial^{-1}q)\\
\qquad\quad+\frac{1}{m}\left(-p\partial^{-1}q-(q^T\partial^{-1}p^T)^T-(m\beta r-\sum\limits_{i=1}^{m}q_i\partial^{-1}p_i)I_m-2m\beta(\partial q^T\partial^{-1}q-m\beta rp\partial^{-1}q)\right).\end{array}\right.
\end{equation*}
The soliton hierarchy \eqref{SH2} also has bi-Hamiltonian structures
\begin{align}
u_{t_n}=K_n=J\frac{\delta \mathcal{H}_{n+1}}{\delta u}=M\frac{\delta \mathcal{H}_{n}}{\delta u},
\end{align}
where the second Hamiltonian operator $M$ is given by $M=\Phi J$. The entries of $M$ are defined by
\begin{equation*}\left\{\begin{array}{l}
M_{11}=(\partial+\sum\limits_{i=1}^m p_i\partial^{-1}p_i)I_m+q^T\partial^{-1}q-(p\partial^{-1}p^T)^T+2m\beta(\partial p\partial^{-1}q-q^T\partial^{-1}p^T\partial)\\
\qquad\quad+2m^2\beta^2(rq^T\partial^{-1}q+q^T\partial^{-1}qr)-4m^2\beta^2(q^T\partial^{-1}q\partial q^T\partial^{-1}q+q^T\partial^{-1}p^T\partial p\partial^{-1}q),\\
M_{12}=(m\beta r+\sum\limits_{i=1}^mp_i\partial^{-1}q_i)I_m-q^T\partial^{-1}p^T-(p\partial^{-1}q)^T-2m\beta(\partial p\partial^{-1}p^T+q^T\partial^{-1}q\partial)\\
\qquad\quad-2m^2\beta^2(rq^T\partial^{-1}p^T+q^T\partial^{-1}p^Tr)+4m^2\beta^2(q^T\partial^{-1}p^T\partial p\partial^{-1}p^T+q^T\partial^{-1}q\partial q^T\partial^{-1}p^T),\\
M_{21}=(\sum\limits_{i=1}^m q_i\partial^{-1}p_i-m\beta r)I_m-p\partial^{-1}q-(q^T\partial^{-1}p^T)^T+2m\beta(\partial q^T\partial^{-1}q+p\partial^{-1}p^T\partial)\\
\qquad\quad-2m^2\beta^2 (p\partial^{-1}qr+rp\partial^{-1}q)+4m^2\beta^2(p\partial^{-1}q\partial q^T\partial^{-1}q+p\partial^{-1}p^T\partial p\partial^{-1}q),\\
M_{22}=(\partial+\sum\limits_{i=1}^m q_i\partial^{-1}q_i)I_m+p\partial^{-1}p^T-(q^T\partial^{-1}q)^T+2m\beta(p\partial^{-1}q\partial-\partial q^T\partial^{-1}p^T)\\
\qquad\quad+2m^2\beta^2(p\partial^{-1}p^Tr+rp\partial^{-1}p^T)-4m^2\beta^2(p\partial^{-1}p^T\partial p\partial^{-1}p^T+p\partial^{-1}q\partial q^T\partial^{-1}p^T).\end{array}\right.
\end{equation*}

%
%
%%\subsection{non-isospectral problem}
%
%\section{generalized KN soliton hierarchy based on sl(n,r)}
%\subsection{isospectral problem}
%\begin{align}
%U=U(u,\lambda )=\begin{pmatrix}-m(\lambda^2+\beta(p^Tp+qq^T))&\lambda q\\
%\lambda p&(\lambda^2+\beta (p^Tp+qq^T))I_m
%\end{pmatrix}.
%\end{align}
%\begin{align}
%W=\begin{pmatrix}a&b\\c&d\end{pmatrix}
%\end{align}
%%\subsection{non-isospectral problem}
%
%
%\section{generalized KN soliton hierarchy based on so(n,r)}
%\subsection{isospectral problem}
%\begin{align}
%U=U(u,\lambda )=\begin{pmatrix}0&-\lambda q&m[-\lambda^2-\beta(p^Tp+qq^T)]\\
%\lambda q^T&0&-\lambda p\\
%m[\lambda^2+\beta(p^Tp+qq^T)]&\lambda p^T&0
%\end{pmatrix}.
%\end{align}
%%\subsection{non-isospectral problem}
%Let us first begin with solving the stationary zero curvature equation
%\begin{align}
%W_x=[U,W], \ W=\begin{pmatrix}0&-c&-a\\c^T&d&-b^T\\a&b&0\end{pmatrix}
%%=\sum\limits_{i\ge 0}
%%\begin{pmatrix}0&-c_i&-a_i\\c_i&0&-b_i\\a_i&b_i&0\end{pmatrix}\lambda^{-i},
%\end{align}
%which gives

The soliton hierarchy \eqref{SH2} is also Liouville integrable, upon noticing that the vector fields $K_n,\,\, n\ge 1$ possess distinct differential orders and that the common conserved functionals $\{\mathcal{H}_n\}_{n=0}^{\infty}$ and symmetries $\{\mathcal{K}_n\}_{n=0}^{\infty}$ commute:
\begin{equation}\left\{\begin{array}{l}
\{\mathcal{H}_l,\mathcal{H}_m\}_{J}=\int ( \frac{\delta\mathcal{H}_l}{\delta u})^T J\frac{\delta \mathcal{H}_m}{\delta u}dx=0,\\
\{\mathcal{H}_l,\mathcal{H}_m\}_{M}=\int ( \frac{\delta\mathcal{H}_l}{\delta u})^T M\frac{\delta \mathcal{H}_m}{\delta u}dx=0,\end{array}\right. \, m\ge 0,
\end{equation}
and
\begin{align}
[K_l,K_m]=K_l'(u)[K_m]-K_m'(u)[K_l]=0,\ l,m\ge 0.
\end{align}

\begin{remark}
By setting $\beta=0$ and $m=1$ throughout Section 3, one can recover results presented by Ma etc. in \cite{MA1}.
\end{remark}
\begin{remark}
By setting $m=1$ throughout Section 3, one can recover the results presented by Li etc. in \cite{LSMY}.
\end{remark}

\section{Conclusions}
In this paper, we propose two generalized multicomponent AKNS-type spectral problems associated with the Lie algebra $sl(m+1,\mathbb{R})$ and anti-symmetric square matrices of $(m+1)$-th order, respectively. We derive their corresponding soliton hierarchies by the standard procedure, establish their bi-Hamiltonian structures by using the trace identity and thus prove the Liouville integrability for the systems in each of the two soliton hierarchies separately. We would like to remark that these two matrix spectral problems and their corresponding results can be reduced to the ones considered in \cite{AKNS},\cite{YZY},\cite{MA6}, \cite{MA1} and \cite{LSMY}, respectively, by choosing $m$ and $\beta$ appropriately. As a side product, we obtain some interesting identities presented in Appendix.

\section*{Acknowledgement}
This work was supported by the National Natural Science Foundation of China under the grants 11271266, 11271008, 11371323, 11371326 and 61072147, Natural Science Foundation of Shanghai (Grant No. 11ZR1414100), Zhejiang Innovation Project of China (Grant No. T200905), and the First-class Discipline of Universities in Shanghai and Shanghai Univ. Leading Academic Discipline Project (No. A.13-0101-12-004).

\appendix
\section*{Appendix A. Identities used in Section 2}
By assuming that both $c^{(k)}$ and $q=(q_1,\cdots,q_m)$ are row vectors of $m$-th order and both $b^{(k)}$ and $p=(p_1,\cdots,p_m)^T$ are column vectors of $m$-th order, it is easy to prove the following identities
\begin{align*}
&\partial^{-1}(c^{(k)}q)p=\sum\limits_{j=1}^m p_j\partial^{-1}q_jc^{(k)},\qquad\qquad\qquad\qquad p\partial^{-1}(b^{(k)}p)=p\partial^{-1}(p^T{b^{(k)}}^T),\\
&\partial^{-1}(pb^{(k)})p={(p\partial^{-1}p^T)}^T{b^{(k)}}^T,\qquad\qquad\qquad\qquad\partial^{-1}({b^{(k)}}^Tp^T)q^T=\sum\limits_{j=1}^mq_j\partial^{-1}p_j{b^{(k)}}^T,\\
&q^T\partial^{-1}({c^{(k)}}^Tq^T)=q^T\partial^{-1}(qc^{(k)}),\,\,\qquad\qquad\quad\qquad\partial^{-1}(q^T{c^{(k)}}^T)q^T=(q^T\partial^{-1}q)^Tc^{(k)},\\
&(\sum\limits_{j=1}^mq_j\partial^{-1}p_j)q^T\partial^{-1}p^T=(q^T\partial^{-1}q)^Tp\partial^{-1}p^T,\qquad(\sum\limits_{j=1}^mq_j\partial^{-1}p_j)q^T\partial^{-1}q=(q^T\partial^{-1}q)^Tp\partial^{-1}q,\\
&(\sum\limits_{j=1}^mp_j\partial^{-1}q_j)p\partial^{-1}p^T=(p\partial^{-1}p^T)^Tq^T\partial^{-1}p^T,\qquad(\sum\limits_{j=1}^mp_j\partial^{-1}q_j)p\partial^{-1}q=(p\partial^{-1}p^T)^Tq^T\partial^{-1}q,\\
&q^T\partial^{-1}p^T(\sum\limits_{j=1}^mq_j\partial^{-1}p_j)=q^T\partial^{-1}q(p\partial^{-1}p^T)^T,\qquad q^T\partial^{-1}q(\sum\limits_{j=1}^mp_j\partial^{-1}q_j)=q^T\partial^{-1}p^T(q^T\partial^{-1}q)^T,\\
&p\partial^{-1}p^T(\sum\limits_{j=1}^mq_j\partial^{-1}p_j)=p\partial^{-1}q(p\partial^{-1}p^T)^T,\,\quad\qquad p\partial^{-1}q(\sum\limits_{j=1}^mp_j\partial^{-1}q_j)=p\partial^{-1}p^T(q^T\partial^{-1}q)^T.
\end{align*}

\section*{Appendix B. Identities used in Section 3}
By assuming that both $b^{(k)}$ and $q=(q_1,\cdots,q_m)$ are row vectors of $m$-th order and both $c^{(k)}$ and $p=(p_1,\cdots,p_m)^T$ are column vectors of $m$-th order, it is easy to prove the following identities
\begin{align*}
&\partial^{-1}(q^Tc^{(k)})q^T=(q^T\partial^{-1}q)^T{c^{(k)}}^T,\qquad\qquad\qquad\qquad\partial^{-1}(p{b^{(k)}}^T)q^T=(q^T\partial^{-1}p^T)^Tb^{(k)},\\
&\partial^{-1}(q^Tc^{(k)})p=(p\partial^{-1}q)^T{c^{(k)}}^T,\quad\qquad\qquad\qquad\qquad\partial^{-1}(p{b^{(k)}}^T)p=(p\partial^{-1}p^T)^Tb^{(k)},\\
&\partial^{-1}({c^{(k)}}^Tq)q^T=\sum\limits_{j=1}^mq_j\partial^{-1}q_j{c^{(k)}}^T,\,\,\qquad\qquad\qquad\quad\partial^{-1}({c^{(k)}}^Tq)p=\sum\limits_{j=1}^mp_j\partial^{-1}q_j{c^{(k)}}^T,\\
&\partial^{-1}(b^{(k)}p^T)q^T=\sum\limits_{j=1}^mq_j\partial^{-1}p_jb^{(k)},\qquad\qquad\qquad\qquad\partial^{-1}(b^{(k)}p^T)p=\sum\limits_{j=1}^mp_j\partial^{-1}p_jb^{(k)},\\
&(\sum\limits_{j=1}^mp_j\partial^{-1}p_j)p\partial^{-1}p^T=(p\partial^{-1}p^T)^Tp\partial^{-1}p^T,\quad\qquad(\sum\limits_{j=1}^mq_j\partial^{-1}q_j)q^T\partial^{-1}q=(q^T\partial^{-1}q)^Tq^T\partial^{-1}q,\\
&(\sum\limits_{j=1}^mp_j\partial^{-1}q_j)q^T\partial^{-1}q=(p\partial^{-1}q)^Tq^T\partial^{-1}q,\qquad\qquad(\sum\limits_{j=1}^mq_j\partial^{-1}p_j)p\partial^{-1}p^T=(q^T\partial^{-1}p^T)^Tp\partial^{-1}p^T,\\
&(\sum\limits_{j=1}^mp_j\partial^{-1}p_j)p\partial^{-1}q=(p\partial^{-1}p^T)^Tp\partial^{-1}q,\,\qquad\qquad(\sum\limits_{j=1}^mq_j\partial^{-1}q_j)q^T\partial^{-1}p^T=(q^T\partial^{-1}q)^Tq^T\partial^{-1}p^T,\\
&(\sum\limits_{j=1}^mq_j\partial^{-1}p_j)p\partial^{-1}q=(q^T\partial^{-1}p^T)^Tp\partial^{-1}q,\qquad\qquad(\sum\limits_{j=1}^mp_j\partial^{-1}q_j)q^T\partial^{-1}p^T=(p\partial^{-1}q)^Tq^T\partial^{-1}p^T,\\
&q^T\partial^{-1}q(\sum\limits_{j=1}^mq_j\partial^{-1}q_j)=q^T\partial^{-1}q(q^T\partial^{-1}q)^T,\,\,\quad\qquad q^T\partial^{-1}p^T(\sum\limits_{j=1}^mp_j\partial^{-1}q_j)=q^T\partial^{-1}p^T(p\partial^{-1}q)^T,\\
&q^T\partial^{-1}p^T(\sum\limits_{j=1}^mp_j\partial^{-1}p_j)=q^T\partial^{-1}p^T(p\partial^{-1}p^T)^T,\qquad q^T\partial^{-1}q(\sum\limits_{j=1}^mq_j\partial^{-1}p_j)=q^T\partial^{-1}q(q^T\partial^{-1}p^T)^T,\\
&p\partial^{-1}p^T(\sum\limits_{j=1}^mp_j\partial^{-1}q_j)=p\partial^{-1}p^T(p\partial^{-1}q)^T,\qquad\qquad p\partial^{-1}q(\sum\limits_{j=1}^mq_j\partial^{-1}q_j)=p\partial^{-1}q(q^T\partial^{-1}q)^T,\\
&p\partial^{-1}p^T(\sum\limits_{j=1}^mp_j\partial^{-1}p_j)=p\partial^{-1}p^T(p\partial^{-1}p^T)^T,\quad\qquad p\partial^{-1}q(\sum\limits_{j=1}^mq_j\partial^{-1}p_j)=p\partial^{-1}q(q^T\partial^{-1}p^T)^T.
\end{align*}

\end{document}